\documentclass[twocolumn,prl,superscriptaddress,showpacs,floatfix]{revtex4}

\usepackage{epsfig,color}
\usepackage{graphicx}
\usepackage{dcolumn}
\usepackage{bm}
\usepackage{setspace}

\begin{document}

\title{Strongly Enhanced Hole-Phonon Coupling in the Metallic State 
of the Dilute Two-Dimensional Hole Gas}
\author{X. P. A. Gao}
\email{xuangao@lanl.gov}
\affiliation{Los Alamos National Laboratory, Los Alamos, NM 87545}
\author{G. S. Boebinger}
\affiliation{National High Magnetic Field Laboratory, Florida State University, Tallahassee, FL 32312}
\author{A. P. Mills Jr.}
\affiliation{Physics Department, University of California, Riverside, CA 92521}
\author{A. P. Ramirez, L. N. Pfeiffer, K. W. West}
\affiliation{Bell Laboratories, Lucent Technologies, Murray Hill, NJ 07974}

\date{\today}

\begin{abstract}
We have studied the temperature dependent phonon emission rate $P$($T$) of 
a strongly interacting ($r_s\geq$22) dilute 2D GaAs hole system using a standard 
carrier heating technique. In the still poorly understood metallic state, 
we observe that $P$($T$) changes from $P$($T$)$\sim T^5$ to $P$($T$)$\sim T^7$ above 
100mK, indicating a crossover from screened piezoelectric(PZ) coupling to screened 
deformation 
potential(DP) coupling for hole-phonon scattering. Quantitative comparison with 
theory shows that the long range PZ coupling between holes and phonons has the 
expected magnitude; however, in the metallic state, the short range DP coupling 
between holes and phonons is {\it almost twenty times stronger} than expected from theory. 
The density dependence of $P$($T$) shows that it is {\it easier} to cool low density 2D holes 
in GaAs than higher density  2D hole systems.   

\end{abstract}
\maketitle

Hot carrier effects in semiconductors have been a subject of interest for device 
applications as well as for probing electron/hole-phonon interactions\cite{ridley}. In the 
low temperature Bloch-Gruneisen regime, hot carrier effects are particularly useful 
for studying inelastic electron/hole-phonon scattering, because direct transport 
measurements require extremely high sample mobility\cite{stormer}. Price provided the first 
theoretical study of hot two-dimensional (2D) electrons in high mobility GaAs 
heterostructures, considering the phonon emissions via DP and PZ coupling\cite{price}. 
The theory was extended further by other authors [4,5a], and has found excellent 
support from various heating experiments on 2D electrons with modest densities 
($\sim 10^{11}$ cm$^{-2}$)\cite{ma,chow,wennberg,appleyard,fletcher}.   

The carrier heating phenomena in high mobility 2D systems with very low carrier densities
has received recent attention\cite{Altshuler,prus}, particularly in the context of the 2D metallic 
transport and the metal-insulator transition (MIT)\cite{mitreview}. As first discovered by Kravchenko 
and coworkers in silicon metal-oxide-semiconductor field-effect-transistors (Si-MOSFET's), 
the apparent metallic-like resistivity of various low density 2D systems challenges the 
celebrated non-interacting scaling theory of localization for Fermi liquids\cite{mitreview}. 
However, because the $T$=0 conductivity determines if the ground state of the system is 
metallic, it is critical to study the metallic-like behavior at the lowest possible 
temperatures. In the low temperature regime, the electron-phonon scattering time becomes 
very long and self-heating of carriers in a driving electric field poses a serious problem
in interpreting experimental data\cite{Altshuler}. To understand the carrier-phonon coupling of 
dilute 2D systems in the anomalous 2D metallic state, in this Letter we present a study 
of the energy relaxation of a dilute metallic 2D hole system (2DHS) in response to a 
driving electric field down to very low temperatures ($\sim$27mK). Note that it is not 
{\it a priori} known if the standard weak-interaction theories of hot carriers
[3, 4, 5a] would also apply to the anomalous 2D metallic state, since the strongly 
correlated 2D metal may not be a Fermi liquid state\cite{mitreview}. 

Our energy relaxation measurements suggest that the dilute metallic 2D holes in our 
GaAs quantum well (QW) is cooled by emitting phonons. For all the densities studied, 
the $T$ dependent phonon emission rate $P$($T$) exhibits either a $T^5$ or $T^7$ power law, with 
a crossover at $\sim$100mK. The $T^5$ and $T^7$ power laws of $P$($T$) are interpreted as arising 
from the screened PZ or DP coupling (respectively) for hole-phonon coupling.  The crossover 
at 100mK suggests the two mechanisms have comparable magnitude at that temperature. 
Comparing our data with existing theories, we find that the $T^5$ part of $P$($T$) below 100mK 
shows excellent quantitative agreement with weak-interaction theories. The $T^7$ part 
of $P$($T$), however, is about {\it three hundred} times larger in magnitude than the expected 
value for normal DP hole-phonon coupling. This indicates that the short range DP coupling 
strength between 2D holes and phonons is enhanced by almost a factor of twenty in the 
anomalous 2D metallic state. It follows that high mobility dilute 2D holes in GaAs are 
cooled via DP- instead of PZ-coupled phonon scattering at temperatures down to 100mK, 
contrary to conventional experience drawn from higher density samples\cite{price,ma,chow,wennberg,appleyard,fletcher}. 
We suggest many-body effects as a possible origin of this more than an order of magnitude 
enhancement of hole-phonon coupling in the dilute limit.   

Our experiments were performed on a high mobility low-density 2DHS in a 10nm wide 
GaAs QW similar to those in previous studies\cite{Gao}. The sample was symmetrically 
doped with a density of 1.3$\times$10$^{10}$cm$^{-2}$ from doping. The hole density $p$ 
was tuned by a backgate about 0.15mm below the QW. The ungated sample has a 
low temperature hole mobility, $\mu\approx 5\times 10^5$cm$^2$/Vs. The sample was prepared in the 
form of Hall bar, with an approximate total sample area 0.2cm$^2$. With the relatively large sample 
area, four-wire transport measurement can be realized with dissipation power down to 
fWatts/cm$^2$ for sample resistance in the k$\Omega$ and above range. The sample was 
driven with 8Hz square wave voltage ranging from $\mu$Vs to mVs in amplitude.  
During the experiments, the sample was immersed in the $^3$He/$^4$He mixture in a top-loading 
dilution refrigerator\cite{footnote13}.
\begin{figure}[btph]
\centerline{\psfig{file=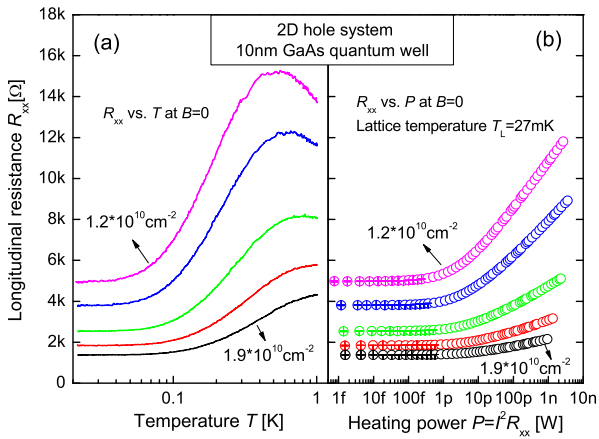,width=9cm}}
\caption{(color online) (a) Longitudinal resistance $R_{xx}$ vs. $T$ at zero magnetic field 
of 2D holes in a 10nm wide GaAs quantum well. The data were obtained with less than 5fWatts 
heating power on the sample. (b) Variation of $R_{xx}$ against heating power $P$=$I^2R_{xx}$ with 
lattice held at $T_L$=27mK. The densities are 1.2, 1.3, 1.5, 1.7 and 1.9$\times$10$^{10}$cm$^{-2}$ 
from top to bottom in both (a) and (b). }
\label{fig1}
\end{figure}

\begin{figure}[hbtp]
\centerline{\psfig{file=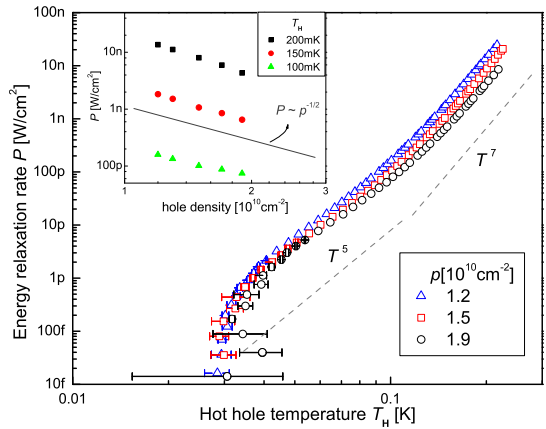,width=8.5cm}} 
\caption{(color online) The energy relaxation rate $P$ vs. hot hole temperature $T_H$ with 
the lattice held at $T_L$=27mK, created from data in Fig.1b using $R_{xx}$($T$) in Fig.1a 
as thermometer for hot holes. The dashed lines represent $T^5$ and $T^7$ behavior to be 
compared with the data. The inset plots the power per cm$^2$ needed to heat up holes to 100, 150 and 
200mK with $T_L$=27mK vs. hole density $p$, which is consistent with a $1/\sqrt{p}$ dependence.  
}
\label{fig2}
\end{figure}

We studied heating effects for five different hole densities in the sample from 1.2 to 
1.9$\times$10$^{10}$cm$^{-2}$ deep in the metallic state where the low $T$ resistivity 
is less than 0.1h/e$^2$ for all the densities. Using an effective hole mass $m^*$=0.36$m_e$, 
we obtain 22$\leq r_s\leq$28 for this density range, where $r_s$ is the ratio between the 
inter-particle Coulomb energy and the Fermi energy. During the experiment and following 
previous experiments on electron-phonon coupling\cite{wennberg,fletcher,prus} the sample 
resistance was used as a self-thermometer for the overheated holes. In Fig.\ref{fig1}a, 
the $T$ dependence of $R_{xx}$, the longitudinal resistance is shown on a semi-log plot, 
where the resistivity $\rho_{xx}$ would be roughly $R_{xx}$ divided by three.    
The data in Fig.\ref{fig1}a were obtained with negligible Joule heating power ($<$5fWatts) 
on the sample to ensure good thermal equilibrium between holes and the lattice. 
The 2D holes were then purposely overheated by increasing the current through sample, 
with the lattice/substrate being held at a fixed temperature $T_L$. Figure1b shows the 
resulting Joule heating power $P$=$I^2R_{xx}$ dependence of the sample resistance $R_{xx}$ 
for $T_L$=27mK.  Except at very low power, the resistance increases as the power is increased, 
indicating the hole gas being heated up. Using the $R_{xx}$($T$) curve in Fig.\ref{fig1}a 
as a thermometer for the hot holes, one can deduce the hot hole temperature $T_H$ for 
given $R_{xx}$ and power $P$ in Fig.\ref{fig1}b. In this way, we obtain $P$ 
(i.e. energy relaxation rate) vs. $T_H$ from the $R_{xx}$($P$) data in Fig.\ref{fig1}b, 
as shown in the main panel of Fig.\ref{fig2}. For the clarity of presentation, we only 
include data for three densities.

We discuss some important qualitative features in the energy relaxation rate $P$ vs. 
hot hole temperature $T_H$ data of Fig.\ref{fig2}.  First, $P$($T$) follows $T^5$ 
dependence and crosses over to a $T^7$ dependence above $\sim$100mK. The exponent of 
the power law dependence of $P$($T$) can be used to distinguish the type of carrier-phonon 
coupling. If the carriers emit acoustic phonons via screened PZ or DP coupling, then the energy 
relaxation $P$($T$) will have a $T^5$ or $T^7$ dependence. The $T^5$ dependence we observe 
below 100mK is consistent with the acoustic phonon emission via screened PZ coupling, 
similar to previous theories and experiments on 2D electrons in GaAs\cite{price,ma,chow,
wennberg,appleyard,fletcher}. The $T^7$ suggests the dominance of screened DP coupling 
for phonon emission above 100mK for dilute 2D holes.  Interestingly, this crossover in 
power law dependence of $P$($T$) was predicted by theory to happen around 2K for long 
time\cite{price}, but has never been observed experimentally. We shall see that an order 
of magnitude enhancement in the DP coupling strength can move the crossover temperature 
down to $\sim$100mK, as in our case. Secondly, the $P$($T$) curves for different densities 
show that it takes more power to warm up holes with lower density to a given temperature, 
i.e. it is easier to cool 2D holes with lower density. This is illustrated in the inset of 
Fig.\ref{fig2}, in which we plot the power/cm$^2$ needed to warm holes from 27mK to $T_H$=100, 150 
or 200mK against the hole density. Assuming the formalism of hot electrons theories in 
ref.3 and 4 is also applicable to holes with appropriate substitution of parameters 
(carrier effective mass, DP constant etc.), the energy relaxation rate through acoustic 
phonon emission for clean 2D holes in GaAs is\cite{footnote14}
\begin{equation}
\label{eq1}
P=[15\times (T_H^5-T_L^5)+0.06\times D^2\times (T_H^7-T_L^7)]/\sqrt{p}
\end{equation}
On the right hand side of Eq.\ref{eq1}, the first/second part is the acoustic phonon 
emission rate through screened PZ/DP coupling. In Eq.\ref{eq1}, $P$ is in units of 
$\mu$Watts/cm$^2$, $T_H$ and $T_L$ are in units of Kelvin, hole density $p$ is in units 
of 10$^{10}$cm$^{-2}$, and the DP constant $D$ is in units of eV. Eq.\ref{eq1} predicts 
that the energy relaxation rate is inversely proportional to the square root of density, 
and therefore lower density holes can sustain higher heating power. This result is simply 
related to the fact that the Fermi degeneracy restricts the phonon scattering probability 
and the Fermi temperature is lower in more dilute 2DHS. In the inset of Fig.\ref{fig2} we 
draw a gray line representing the $P\propto 1/\sqrt{p}$ dependence, which is consistent 
with the data.

\begin{figure}[htbp]
\centerline{\psfig{file=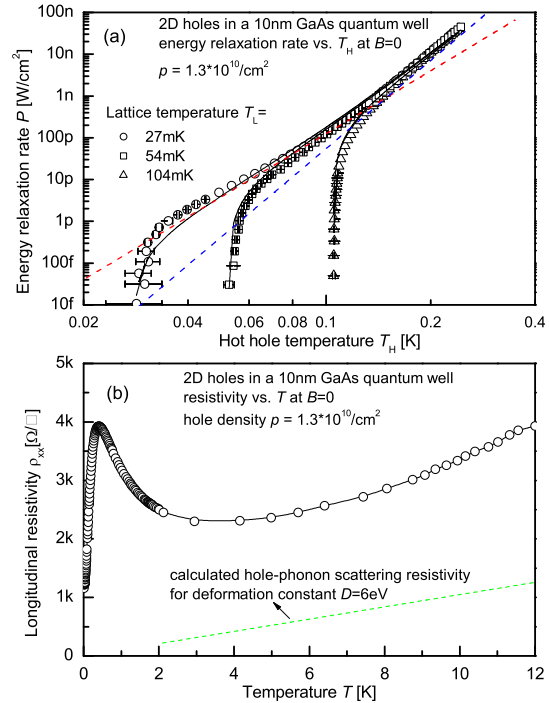,width=8.5cm}}
\caption{(color online) (a) The temperature dependent energy relaxation rate of 2D 
holes with density $p$=1.3$\times$10$^{10}$cm$^{-2}$. Data are shown for three different 
lattice temperatures, $T_L$=27, 54 and 104mK. The black lines are according to Eq.1 with 
an enhanced deformation potential constant $D$=105eV. The red/blue dotted lines are 
respectively the PZ coupling contribution (15$\times T_H^5/\sqrt{p}$) and DP coupling 
contribution (0.06$\times$105eV$^2\times T_H^7$/$\sqrt{p}$) of the energy relaxation rate. 
(b)Resistivity $\rho_{xx}$($T$) over an extended temperature range for the 
2D holes in (a) at zero magnetic field. The dotted green line depicts the hole-phonon 
scattering induced resistivity in the high temperature equi-partition regime with 
deformation potential constant $D$=6eV according to \cite{walukiewicz,karpus}. After
adding a resistivity offset induced by impurity scattering, the hole-phonon scattering
for $D$=6eV in the equi-partition regime accounts for the $T>$4K experimental data satisfactorily.}
\label{fig3}
\end{figure}

Now we compare the energy relaxation rate $P$($T$) data with existing theories quantitatively. 
The best estimate of the deformation potential constant $D$ for hole-phonon scattering in GaAs 
is around 6eV\cite{holeDP}. With $D$=6eV, Eq.\ref{eq1} predicts that below 2.6K the holes mostly 
relax by emitting acoustic phonons via screened PZ coupling. However, our data suggest that it 
is the DP coupled phonon emission that is responsible for the energy relaxation above 100mK. 
We find that it is possible to fit our $P$($T$) data to Eq.\ref{eq1} with $D$ as the only 
fitting parameter. The best fitted $D$ equals 105eV for all the densities studied between 
1.2 and 1.9$\times$10$^{10}$cm$^{-2}$. In other words, standard theories of hot carriers in 
the Bloch-Gruneisen regime can explain our data with an {\it unaltered} PZ coupling and an 
{\it eighteen times enhanced} DP coupling. In Fig.\ref{fig3}a, we plot the $P$($T_H$) for density 
1.3$\times$10$^{10}$cm$^{-2}$ at three lattice temperatures: 27, 54 and 104mK. The black 
lines are the theoretical curves according to Eq.\ref{eq1} with $D$=105eV, which have good 
agreement with data. In Fig.\ref{fig3}a, we also draw red and blue dotted lines to show the 
absolute (with $T_L$=0) magnitudes of PZ and DP contributions to energy relaxation rate.    
   
The anomalously large deformational potential constant of 105eV fitted from our energy 
relaxation rate data is most likely due to other hole-phonon coupling mechanisms in the 
dilute regime, instead of an unrealistically large DP constant. In contrast to 
electrons\cite{stormer}, the mobility of the 2DHS is not high enough to allow separating 
of phonon scattering from impurity scattering to deduce the DP constant $D$ in the 
Bloch-Gruneisen regime.  However, it is possible to estimate $D$ from the phonon-scattering 
induced resistivity (denoted as $\rho_{ph}$($T$) hereafter) in the high temperature 
equi-partition regime\cite{walukiewicz,karpus}. In Fig.\ref{fig3}b we show the 
resistivity $\rho_{xx}$($T$) at $B$=0 over an extended temperature range up to 12K for the
2DHS with $p$=1.3$\times$10$^{10}$cm$^{-2}$. Above 4K, the resistivity has a positive 
temperature coefficient caused by phonon-scattering in the equi-partition regime. We also 
calculated $\rho_{ph}$($T$) with $D$=6eV in this regime according to ref.\cite{walukiewicz,karpus}
and included the result as a green dotted line in Fig.\ref{fig3}b. It is known that in the high $T$ 
equi-partition regime, DP coupling accounts for most ($\sim$80$\%$) of 
$\rho_{ph}$($T$)\cite{walukiewicz,karpus}. 
Note that 6eV for the DP constant gives a slope that is consistent with the observed slope of 
$\rho_{xx}$($T$). 
If the DP constant were 105eV, the theoretical curve for $\rho_{ph}$($T$) 
would be way out of scale in Fig.\ref{fig3}b, as $\rho_{ph}$($T$)$\propto D^2$ \cite{walukiewicz,karpus}. 
We also briefly comment on the non-monotonic peak around $T$=0.5K of $\rho_{xx}$($T$) at lower 
temperatures. This first increasing and then decreasing resistivity of dilute 2D systems with 
reducing temperature was attributed to the degeneracy of the systems in which the Fermi 
temperatures are very low\cite{DasSarma}. To be consistent with this classical-quantum crossover 
scenario, we expect to observe the energy relaxation rate $P$($T$) changes from $T^5$(or $T^7$) 
into a linear $T$ dependence for $T>$0.5K \cite{price}. Unfortunately, our sample could not sustain larger 
current to check this power law crossover in $P$($T$) around 0.5K. 

Jain, Jalabert and Das Sarma(JJD) proposed a mechanism for energy relaxation rate enhancement of 
electron gas where new low-energy phonon modes are involved\cite{JJD,DJJ,KDJJ}.  
These new phonon modes are supposed to be due to coupling between quasiparticle excitations 
and longitudinal {\it optical} (LO) phonons. Including the extra energy loss of hot electrons 
through these novel phonon modes, a slightly enhanced effective DP constant of 16eV in 
experiment\cite{manion} could be explained\cite{JJD,DJJ,KDJJ}. Note that the experiments by 
Manion {\it et al.} were performed on 2D electrons with fifty times higher density, and at 
much higher temperatures, relevant to the excitation energy scales in JJD theory,
as oppose to our low $T$ Bloch regime study. Moreover, JJD theory predicts a greatly
enhanced energy relaxation rate in {\it optical} phonon scattering, while our data are
consistent with enhanced {\it acoustic} phonon scattering via DP coupling. 
With the detailed temperature and density dependences to be further 
calculated in the JJD model into the density and temperature ranges of our experiment, 
the relevance between the many-body effects and renormalized LO phonon modes \cite{JJD,DJJ,KDJJ} 
with our data remains to be seen.
 
Finally, we discuss some relevance of our results to the much debated 2D metallic state problem
\cite{Altshuler,mitreview}. First, it is striking that the hot carrier theories for weakly 
interacting systems appear to account for important features (i.e. the $T^{5, 7}$ and the 
1/$\sqrt{p}$ dependences of energy relaxation rate $P$) of our data. Within the two-temperature 
model, the energy relaxation rate is related to the specific heat $C$ of the 2D holes and the 
hole-phonon scattering rate $\tau_{ph}^{-1}$ as $P(T_H)=\int_{T_L}^{T_H}C\times\tau_{ph}^{-1}dT$. 
The power law dependence of $P$($T_H$) we observed is consistent with a linear $T$ dependent 
specific heat and $T^{3, 5}$ dependence of $\tau_{ph}^{-1}$ for our dilute holes in the metallic 
state. Linear $T$ dependent $C$($T$) is a strong indication of Fermi-liquid, in contrast to 
some non-Fermi liquid models of the 2D metallic state where $C$($T$) might have a non-trivial 
$T$ dependence\cite{mitreview}. Next, our measurements provide evidence that the minute energy 
loss through emitting phonons at low $T$ poses a more severe problem on performing experiments 
on samples with higher carrier density, lower resistance and smaller area.
  
The authors acknowledge useful discussions with S.H. Simon.
X.P.A. Gao acknowledges the director's funded postdoctoral fellowship at LANL.  
The NHMFL is supported by the NSF and the State of Florida.


\begin{references}

\bibitem{ridley}B.K. Ridley,{\it Rep. Prog. Phys.} {\bf 54}, 169 (1991).
\bibitem{stormer}H.L. Stormer, L.N. Pfeiffer, K.W. Baldwin, and K.W. West, 
{\it Phys. Rev. B} {\bf 41}, 1278 (1990).
\bibitem{price}P.J. Price, {\it J. Appl. Phys.} {\bf 53}, 6863 (1982).  
\bibitem{ma}Y. Ma {\it et al.}, {\it Phys. Rev. B} {\bf 43}, 9033 (1991).   
\bibitem{chow}(a)E. Chow, H.P. Wei, S.M. Girvin and M. Shayegan, 
{\it Phys. Rev. Lett.} {\bf 77}, 1143 (1996); (b)E. Chow, H.P. Wei, 
S.M. Girvin, W. Jan and J.E. Cunningham, {\it Phys. Rev. B} {\bf 56}, R1676 (1997).   
\bibitem{wennberg}A.K.M. Wennberg {\it et al.}, {\it Phys. Rev. B} {\bf 34}, 4409 (1986).
\bibitem{appleyard}N.J. Appleyard, J.T. Nicholls, M.Y. Simmons, W.R. Tribe and 
M. Pepper, {\it Phys. Rev. Lett.} {\bf 81}, 3491 (1998).         
\bibitem{fletcher} R. Fletcher, Y. Feng, C.T. Foxon and J.J. Harris, 
{\it Phys. Rev. B} {\bf 61}, 2028 (2000).
\bibitem{Altshuler}B.L. Altshuler, D.L. Masolov and V.M. Pudalov, 
{\it Physica E} {\bf 9}, 209 (2001).
\bibitem{prus}O. Prus, M. Reznikov, U. Sivan and V. Pudalov, {\it Phys. Rev. Lett.} 
{\bf 88}, 016801 (2002).                                                                                                                    
                                                                                                                                                                                                                                                                                                             
\bibitem{mitreview}E. Abrahams, S.V. Kravchenko, and M.P. Sarachik, 
{\it Rev. Mod. Phys.} {\bf 73}, 251 (2001).

\bibitem{Gao}X. P.A. Gao, A.P. Mills, Jr., A.P. Ramirez, L.N. Pfeiffer, and K.W. West, 
{\it Phys. Rev. Lett.} {\bf 89}, 016801 (2002); {\it ibid} {\bf 88}, 166803 (2002).

\bibitem{footnote13}In a preliminary study (cond-mat/0203151) of us on a 30nm GaAs QW, 
where the sample was mounted on the copper tail of the mixing chamber of a different 
dilution refrigerator, we also observed a $T^5$ dependence of $P$($T$) with similar magnitude 
below 0.1K in the metallic state.                                                          

\bibitem{footnote14}We obtained our Eq.1 here using the Eqs. A17, A20 and A21 by 
Ma {\it et al.} in ref.4 with a correction factor of 0.77 for the phonon anisotropy as 
shown in C. Jasiukiewicz and V. Karpus, {\it Semicond. Sci. Technol.} {\bf 11}, 1777 (1996).
Interestingly, according to the calculations by Ma {\it et al.}, carrier effective mass 
has no effect in the formula of $P$($T$) if the carrier-phonon interaction is screened. 
Therefore, the prefactors of the $T^5$ and $T^7$ terms in our Eq.1 are the same for both 
electrons and holes in GaAs, with appropriate DP constant $D$.                                                                         
                                                             
\bibitem{holeDP}D.D. Nolte, W. Walukiewicz and E.E. Haller, {\it Phys. Rev. Lett.} {\bf 59}, 501 (1987).
\bibitem{walukiewicz}W. Walukiewicz, {\it J. Appl. Phys.} {\bf 59}, 3577 (1986); 
{\it Phys. Rev. B} {\bf 37}, 8530 (1988).
\bibitem{karpus}V. Karpus, {\it Semicond. Sci. Technol.} {\bf 5}, 691 (1990).                                                                
\bibitem{DasSarma}S. Das Sarma and E. H. Hwang, {\it Phys. Rev. Lett.}
{\bf 83}, 164 (1999); {\it Phys. Rev. B} {\bf 61}, R7838 (2000).
\bibitem{JJD}J.K.Jain, R. Jalabert, and S. Das Sarma, {\it Phys. Rev. Lett.} 
{\bf 60}, 353 (1988).                                                                 
\bibitem{DJJ}S. Das Sarma, J.K. Jain and R. Jalabert, 
{\it Phys. Rev. B} {\bf 37}, 4560 (1988); {\it ibid} {\bf 41}, 3561 (1990).
\bibitem{KDJJ}T. Kawamura, S. Das Sarma, R. Jalabert and J.K. Jain, 
{\it Phys. Rev. B} {\bf 42}, 5407 (1990).                                  
\bibitem{manion}S.J. Manion, M. Artaki, M.A. Emanuel, J.J. Coleman and K. Hess, 
{\it Phys. Rev. B}, {\bf 35}, 9203 (1987).
        
\end{references}
\end{document}